# Spatially Distributed Wettability Characterization in Porous Media

Faisal Aljaberi[a], Hadi Belhaj[a,*], Sajjad Foroughi[b], Mohammed Al Kobaisi[a], Martin J. Blunt[b]

[a] Chemical and Petroleum Engineering Department, Khalifa University of Science and Technology, Abu Dhabi P.O. Box 127788, United Arab Emirates
[b] Department of Earth Science and Engineering, Imperial College London, SW7 2AZ London, United Kingdom

## Abstract

An enhanced geometric algorithm for automated pore-by-pore contact angle measurement from micro-CT images, is presented that achieves superior accuracy compared to existing methods through robust fluid-fluid and solid-fluid interface extrapolation. Using this high-resolution data, we generate spatially distributed contact angle maps that reveal previously hidden wettability heterogeneity. Our analysis of mixed-wet systems demonstrates the severe limitations of averaged metrics: a sample with a mean contact angle of 64.7°—conventionally classified as uniformly weakly water-wet—exhibits 40% of its pore space in the intermediate-wetting regime (70-110°). This heterogeneity explains the presence of minimal surface interfaces and fundamentally different pore-filling mechanisms operating within the same sample. By providing open-source tools for spatially-resolved wettability characterization, this work enables more accurate predictions of multiphase flow behavior in heterogeneous porous materials, essential for optimizing subsurface energy storage and recovery processes.

Corresponding author.
E-mail address: hadi.belhaj@ku.ac.ae (H. Belhaj).

# 1. Introduction

Geoengineering, specifically gas storage in geological formations (Krevor et al., 2023), and oil recovery, are essential for the development of sustainable energy resources. These processes are governed by immiscible multiphase flow in porous media, where displacement and trapping mechanisms are predominantly capillary dominated (Singh et al., 2019). Wettability and complex pore structures control these capillary forces (Irannezhad et al., 2022; Primkulov et al., 2018; Raeini et al., 2022; Zhou et al., 2024) that determine fluid invasion, displacement patterns, and trapping efficiency. In $CO_2$ sequestration applications, the wettability of $CO_2$-brine systems varies significantly with storage conditions, such as pressure and temperature. The rock's surface mineralogical and surface roughness further influences wetting behaviour (Blunt, 2017). In oil-brine systems, in addition to the factors mentioned above, surface wettability can be also altered through contact with organic material, such as crude oil containing carboxylic acids, naphthenic acid or/and asphaltenes (Buckley et al., 1998; Mohammed and Babadagli, 2015; Radke et al., 1992). All the aforementioned factors can occur simultaneously and unevenly, leading to spatial variations in wettability even within cm-scale samples.

Wettability originates from molecular interactions that determine liquid-solid adhesion and liquid-liquid cohesion. These interactions define surface energy, representing the system's energy at interfaces. The difference in surface energies between the lighter fluid-solid ($\sigma_{2s}$) and denser fluid-solid ($\sigma_{1s}$) interfaces is key to describing wettability. High adhesion (low $\sigma_{1s}$) or low cohesion promotes spreading of the denser phase, while high cohesion resists it. Young's equation relates these surface energies to the equilibrium contact angle ($\theta_{eq}$) measured through the denser phase 1:

$$cos\theta_{eq} = \frac{\sigma_{2s} - \sigma_{1s}}{\sigma} \tag{1}$$

Here, $\sigma$ represents the interfacial tension between the two fluids 1 and 2, balancing adhesive and cohesive forces at the molecular level. This equation balances forces while unifying the energy approach (minimizing free energy) and geometric approach (contact angle), providing a convenient way to quantify surface energy across diverse systems (Armstrong et al., 2021; Blunt, 2017).

X-ray micro-computed tomography (micro-CT) now enables micro-scale visualization of porous material samples, providing 3D images of pore space geometry for fluid dynamics simulations. During flow experiments, sequential imaging captures fluid-fluid interfaces at various saturation points, allowing characterization of fluid configurations and interface properties for capillary pressure and wettability quantification. Contact angles can be measured *in situ*, three of these main approaches: direct measurement from 3D images (AlRatrout et al., 2017; Andrew et al., 2014; Prodanovic et al., 2006; Scanziani et al., 2017), energy balance calculations using fluid-fluid interface curvature and changes in saturation and interface areas between saturation points (Blunt et al., 2019), or most recently through the topological principle of the isolated phase clusters (Sun et al., 2020a, 2020b, 2020c; Wang et al., 2024). Each approach has distinct advantages and limitations as outlined in Table 1.

**Table 1.** Summary of key pore-scale contact angle measurement approaches, including their methodology, advantages, limitations, and applicable material contexts.

| Measurement Approach | Key Publications | Methodology Description | Pros | Cons | Material Context |
|---|---|---|---|---|---|
| **Direct micro-CT measurement** ($\theta_g$) | Andrew et al., 2014 (Manual) AlRatrout et al., 2017 (Automated) | Directly measure geometric contact angles from high-resolution microtomographic images at the pore scale. | - Direct, straightforward measurement. <br> - Captures local heterogeneity and hysteresis. <br> - High accuracy in ideal conditions and high-quality images. | - Sensitive to image quality. <br> - Difficulties handling rough and irregular surfaces. <br> - Measure the angle observed from the images: at steady-state these angles at rest not the angle during a displacement. | General porous media with any fluid configuration |
| **Thermodynamic Energy Balance** | Blunt et al., 2019 | Solve for contact angle using thermodynamic principles (Helmholtz free energy), assuming equilibrium states, reversible processes, and no energy loss. | - Simple. <br> - Effective in quantifying average wettability. <br> - Measures effective contact angle during displacement. <br> - Less sensitive to local segmentation issues compared to direct measurement. | - Neglect of viscous and dissipative losses. <br> - Over-simplifies complex wetting dynamics where fluid viscosity effects and flow kinetic energy are present. <br> - Requires multiple images to measure energy changes. <br> - Cannot capture system wetting heterogeneity. | Systems with low external energy (water-wet sandstone rock) with any fluid configuration |
| **Gauss-Bonnet (original topological)** | Sun et al., 2020a, 2020b, 2020c | Utilizes the Gauss-Bonnet theorem, relating interface curvature and topological measures (Euler characteristic), to derive wettability metrics from continuum to pore scales. | - Theoretically robust and innovative providing linkage between continuum and pore-scale angles. <br> - Less sensitive to local segmentation issues compared to direct measurement. | - Requires accurate curvature calculations. <br> - Less reliable in mixed-wet conditions. <br> - Assumes simplified conditions that may overlook complex contact line dynamics. <br> - Cannot capture system wetting heterogeneity. | General porous media with isolated phase clusters |
| **Extended Gauss-Bonnet topological** | Wang et al., 2024(Wang et al., 2024) | Enhances original Gauss-Bonnet method by accounting for surface orientation at the contact line, providing improved | - Improved accuracy across diverse wetting scenarios, including mixed-wet conditions. | - Even with improvements, it still produces results closer to but not superior to geometric contact angles in both ideal and real cases. | General porous media with isolated phase clusters |

| accuracy across diverse wettability conditions. | - Cannot capture system wetting heterogeneity. |

Despite these challenges, *in situ* contact angle measurements remain essential for characterizing displacement behavior in experimental studies (Alhammadi et al., 2020; Higgs et al., 2024; Jangda et al., 2023) and for providing key inputs to pore-scale simulations (Giudici, 2023; Jahanbakhsh et al., 2021; Zhou et al., 2024). For a comprehensive review of wettability assessment techniques across scales, see (Armstrong et al., 2021). While more theoretically elaborate methods exist, the geometric method $\theta_g$ remains superior for characterizing mixed-wet systems (Wang et al., 2024). It uniquely captures spatial variations in wettability across solid surfaces with high accuracy, especially when applied to high-resolution images.

An example highlighting the importance of measuring spatially distributed wettability is its influence on two key macroscopic properties, capillary pressure and relative permeability, that are governed by pore-structure, displacement direction (drainage or imbibition), and wettability (Blunt, 2017). However, even with accurate pore-space geometry representation, pore-scale simulation often shows persistent discrepancies with experimental relative permeability and capillary pressure measurements when using an average contact angle. The spatial distribution of wettability significantly impacts macroscopic multiphase flow parameters, as it controls displacement sequences and fluid morphology (Armstrong et al., 2021). This suggests that using average wettability alone is insufficient to fully characterize multiphase flow behavior. Recent studies have highlighted the importance of spatial wettability distribution in pore-scale simulation: (Foroughi et al., 2020) demonstrated that pore-by-pore wetting distribution improves classical Pore Network Model (PNM) predictions, while (Raeini et al., 2022) used Generalized Network Models (GNM) incorporating complex pore shapes to show how wettability strongly influences displacement efficiency and capillary pressure. Their work revealed that heterogeneous wettability leads to more irregular trapping and flow patterns compared to uniform wettability distributions, with different flow regimes emerging based on variations in wettability and capillary numbers. Further advancing this understanding (Zhou et al., 2024) investigated the combined effects of pore geometry and wettability on two-phase flow dynamics using direct numerical simulations. Their work revealed that wettability significantly influences interfacial curvature and capillary pressure, with intermediate wettability (contact angles ~75°) causing curvature rearrangement or reversal, leading to temporary decreases or even negative capillary pressure in converging pore sections.

In this paper, we present a fully automated algorithm for extracting spatially distributed geometric contact angles from high-resolution voxel-based images. A key challenge lies in achieving accurate pore-by-pore measurements $\theta_g$ near three-phase contact regions, where segmentation artifacts and partial volume effects often introduce uncertainty. To address this, we developed a novel surface reconstruction approach that refines normal vector calculations by incorporating local neighborhood information and explicitly accounts for uncertainties at phase boundaries. This reduces noise in the derived contact angles and limits errors associated with discrete voxel segmentation. Furthermore, our method incorporates physics-driven interpolation to extend wettability characterization into regions where direct measurements are not feasible. Spatial mapping of contact angles reveals details that average measurements may mischaracterize the system.

## 2. Image Data Used and Processing Steps

We assembled two complementary datasets to evaluate our contact-angle measurement code in porous media containing two immiscible phases: (i) synthetic voxelized images with known analytical solutions, and (ii) high-resolution micro-CT scans of rock samples subjected to flooding experiments.

### 2.1. Synthetic Voxelized Data

Two synthetic voxelized test cases were generated to validate the performance of our method (Fig 1). The first test case consists of a spherical droplet resting on an ideal flat plane, spanning contact angles from 30° to 150° to cover the full spectrum from strongly hydrophilic to strongly hydrophobic conditions. This simple geometry admits an exact analytical solution and tests algorithmic accuracy across the entire wettability range. To assess the influence of image resolution on measurement accuracy, we rendered droplets with radii of 50, 28, 14, 10, and 6 voxels. The second test case features a spherical droplet in contact with a spherical grain, yielding a non-planar interface that better reflects the rounded or angular grains encountered in natural porous media. We derived the exact solution for this curved-surface geometry and used it to probe our method's robustness when handling heterogeneous, curved interfaces typical of real geological materials.

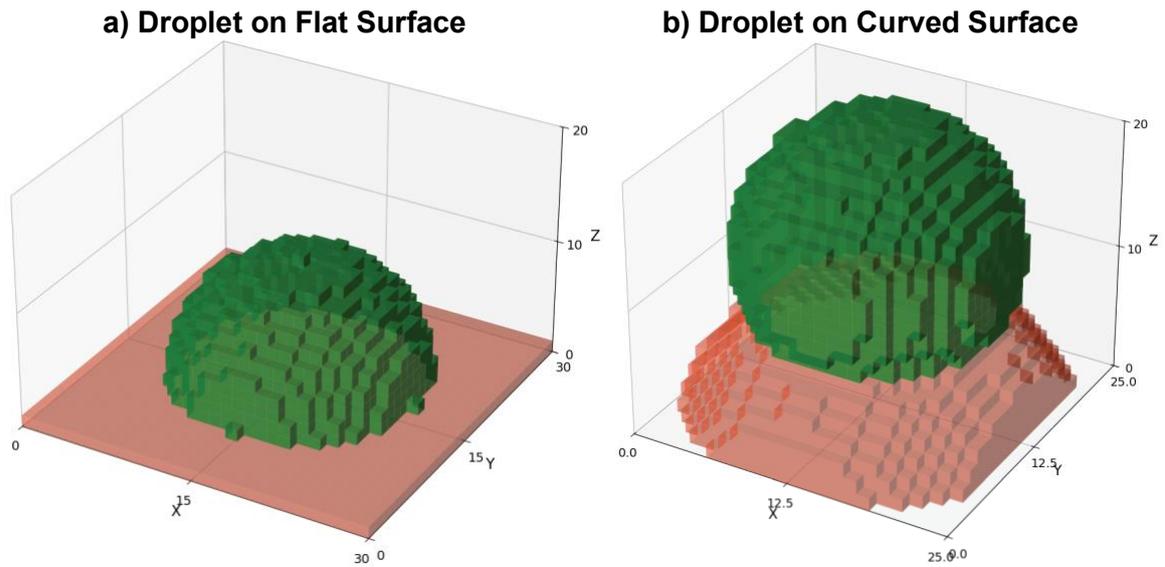

**Fig 1.** Synthetic voxel models of a droplet on (a) a flat surface and (b) a curved surface. Green represents the droplet; red represents the solid. These cases were used to validate the contact angle measurements.

## 2.2. Complex Real Porous Media
### 2.2.1. Raw Data

We further test our code on five high-resolution micro-CT images from flooding experiments, each acquired at voxel sizes small enough to resolve fine interfacial features and capture realistic pore-scale geometries. The raw images of case 1 and 2 were acquired from Digital Porous Media Portal (formally the Digital Rocks Portal).

[1] **Isolated Oil Ganglia Datasets**

   Two high-resolution micro-CT datasets were obtained from a strongly water-wet Ketton limestone sample following an oil injection (primary drainage) and subsequent waterflooding cycle (Scanziani et al., 2017; Singh et al., 2017). These volumes, acquired at 2 μm voxel size, capture isolated oil ganglia trapped within brine-saturated pores,
   - **Ganglion SSb:** A smaller and more geometrically regular ganglion.
   - **Ganglion SSa:** A larger ganglion with more complex morphology.

These datasets were selected to study ganglion shape and trapping behavior under well-controlled wetting conditions.

[2] **Waterflooding Datasets**

   A Bentheimer sandstone sample was initially saturated with brine, followed by oil injection to establish initial conditions. Brine was then injected to initiate oil recovery. A fractional-step waterflooding method was employed, in which brine and oil were sequentially injected in discrete steps with incrementally increasing brine fractions, continuing until steady-state fluid configurations were achieved. At each step, micro-CT scans were acquired at a voxel size of 3.58 μm. Experiments were conducted under two wetting conditions.

- o **Water-wet:** No wettability alteration was applied (Lin et al., 2018).
- o **Mixed-wet:** Wettability was modified through contact with crude oil prior to waterflooding (Lin et al., 2019).

[3] **Hydrogen Storage Dataset**

A recent experiment on a strongly water-wet Bentheimer sandstone investigated hydrogen injection into initially brine-saturated pores (Waleed et al., 2024). Hydrogen formed disconnected ganglia within larger pores, resulting in non-continuous, high-curvature fluid interfaces under multiphase conditions. This dataset provides a new challenge for pore-scale analysis.

## 2.2.2. Imaging process

Although quantitative analysis can be conducted directly from raw images, it suffers from unnecessary complexity—particularly when assessing interfacial properties such as wettability—due to the direct interpretation of greyscale distributions. A widely adopted practice is to segment the image by assigning each phase a unique label. From this, properties such as porosity, morphology, and interfacial characteristics can be extracted more efficiently (Wildenschild and Sheppard, 2013). For automated contact angle measurements, segmenting the image is essential to accurately isolate the interfaces between phases. Prior to segmentation, a noise filter is applied to reduce random-valued impulse noise Fig 2a. The grayscale values in these images originate from projections captured by the scanning system's detector, which are reconstructed into a 3D volumetric image where each voxel's intensity reflects the degree of X-ray attenuation. However, such images are often affected by noise, arising from various sources including limitations of the X-ray detector (e.g., shot noise) and blurry boundaries between phases caused by the partial volume effect (Gonzalez and Woods, 2018; Wildenschild and Sheppard, 2013).

An edge-preserving filtering approach was applied to all raw images. Initially, a non-local means filter (Buades et al., 2005) was utilized to suppress random impulse noise within the bulk phases and to narrow the histogram distribution of each phase, while preserving interface clarity. This filter works by averaging each voxel with its neighboring voxels, weighted according to the similarity between the target voxel and its neighbors. Following this filtering step, segmentation was performed. To enhance the performance of edge detection algorithms— such as the Canny operator, which will be discussed next—edge sharpening was applied using an unsharp masking technique (Gonzalez and Woods, 2018). This process involves three steps: first, the original image is blurred using a Gaussian filter; second, the blurred image is subtracted from the original, resulting in a mask image; and finally, this mask is multiplied by a weighting factor and added back to the original image. From a grayscale intensity signal perspective, when transitioning directionally from low-phase to high-phase intensity, the original image typically shows a continuous gradient (Fig 2b). In contrast, the unsharp masking filter introduces localized oscillations at phase boundaries (Fig 2c), which enhances edge features and improves the response of edge detection algorithms such as Canny. However, segmenting a three-phase system directly from the unsharp-masked image can lead to mislabeling of voxels at phase interfaces, due to the overshoot and undershoot artifacts in grayscale intensity introduced by the sharpening process. Additionally, the sharpening effect can amplify residual noise within the bulk phases that remain after applying the non-local means filter, further complicating accurate segmentation.

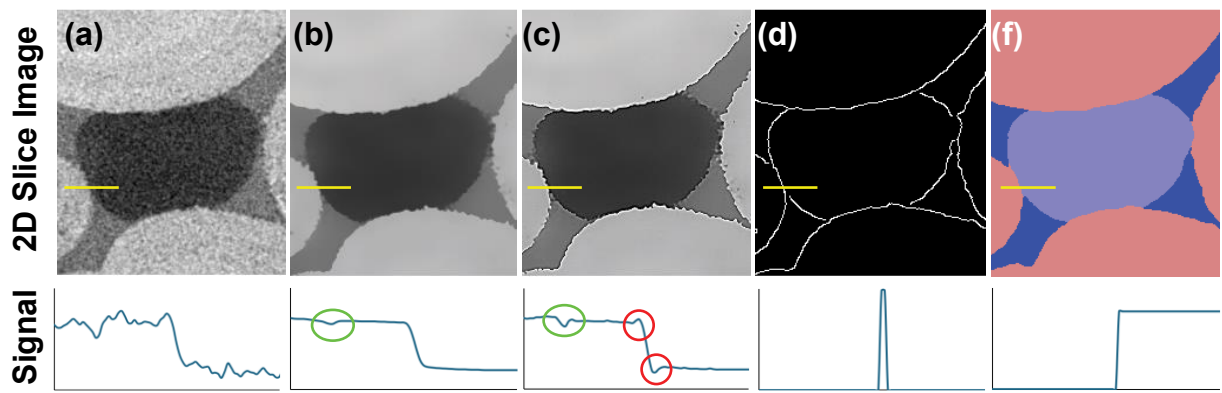

**Fig 2.** Image processing workflow on a 2D slice of the SSb ganglion for contact angle measurement and phase segmentation. The top row shows: (a) raw image, (b) after non-local means (NLM) filtering, (c) after unsharp masking, (d) binary edge-detected image for phase boundary identification, and (f) segmented image (red: solid, purple: brine, blue: oil). The yellow line indicates the location of the 1D gray-value profile shown in the bottom row, illustrating phase transitions from solid (high intensity) to oil (low intensity). The unsharp mask in (c) enhances gradient sharpness, improving phase distinction. Manual measurement follows the sequence (a) → (b) → (c) → (d), while automated segmentation follows (a) → (b) → (f).

The edge image shown in Fig 2d is a binary image where bright regions correspond to interfaces between phases. This image was used for manual contact angle measurements (Section 4.1). The edge detection is based on identifying sharp intensity gradients, which indicate phase changes. Edges are detected using both first and second derivatives of the intensity signal. For accurate 2D interface detection, we employed the Canny edge detection algorithm (Canny, 1986), which is well regarded for its robustness in detecting well-defined boundaries while minimizing artifacts caused by noise.

For the segmentation process, three phases were assigned in all cases: the solid phase, a denser fluid phase (brine), and a lighter fluid phase (oil or gas, depending on the specific image) Fig 2f. The watershed segmentation method was used to segment these phases (Beucher and Meyer, 2018). This method was selected because it is based on a morphological interpretation of the image and divides the segmentation process into two distinct steps. The first step involves defining marker labels, where known grayscale value ranges are assigned to each label—solid, denser fluid, lighter fluid, and unknown. The "unknown" regions typically correspond to boundaries formed where different labeled regions meet. The second step involves resolving the unknown regions using the watershed transform algorithm. The algorithm automatically determines the boundary voxels between phases, resulting in clearly separated regions. In our case, this approach produced highly accurate boundary labeling.

## 3. Measurement of Contact Angle
### 3.1. Manual method for contact angle estimation from 3D images

To manually measure contact angles from real 3D images, we followed the method of Andrew et al., (2014), with some changes. The procedure consists of the following steps.

1. From the segmented image, three-phase contact loops were extracted and smoothed (see Section 4.2).

2. For each node on the contact loop, a secant line was defined based on its neighboring nodes.

3. A 2D grayscale slice, perpendicular to the secant line, was extracted from the raw image.

4. This 2D slice was then processed using the edge-detection steps described in Section 3.2.2. In contrast to the method of Andrew et al., (2014), which used filtered images for fitting interface lines, we used the edge image directly to reduce user subjectivity. The detected edge more accurately represents the true interface, whereas relying on fitted lines from filtered images can introduce significant bias.

5. The contact angle was measured from the resulting edge image by drawing two tangent lines at the three-phase contact point using the open-source software 3D Slicer. These tangent lines were drawn relative to the contact node itself, rather than following the detected edge paths directly, which would otherwise lead to overestimation of the contact angle.

### 3.2. Geometric contact angle measurements

The geometric contact angle ($\theta_g$) is calculated at each node along the three-phase contact loop from

$$\theta_{g,i} = \arccos(n_{fs,i} \cdot n_{ff,i}) \times \frac{180}{\pi} \qquad (2)$$

where $n_{fs,i}$ and $n_{ff,i}$ represent the normal vectors to the fluid-solid interface ($I_{sf}$) and fluid-fluid interface ($I_{ff}$) respectively, as illustrated in Fig 3. The first automated algorithm employing normal vectors from extracted interface meshes was developed by AlRatrout et al. (2017), comprising four main steps: surface extraction, Gaussian smoothing, curvature smoothing, and contact angle calculation.

However, this algorithm presents two significant limitations. First, the curvature smoothing step, which enforces constant curvature on the fluid-fluid interface through iterative vertex adjustment, can substantially deform the interface mesh, with a tendency to shrink the interface, leading to an overestimation of curvature, as demonstrated in Fig 4 where the curvature estimated using commercial image analysis software (Avizo) is compared to the method of AlRatrout *et al.* (2017). Second, the method fails to account for segmentation artifacts that become increasingly pronounced near three-phase contact regions.

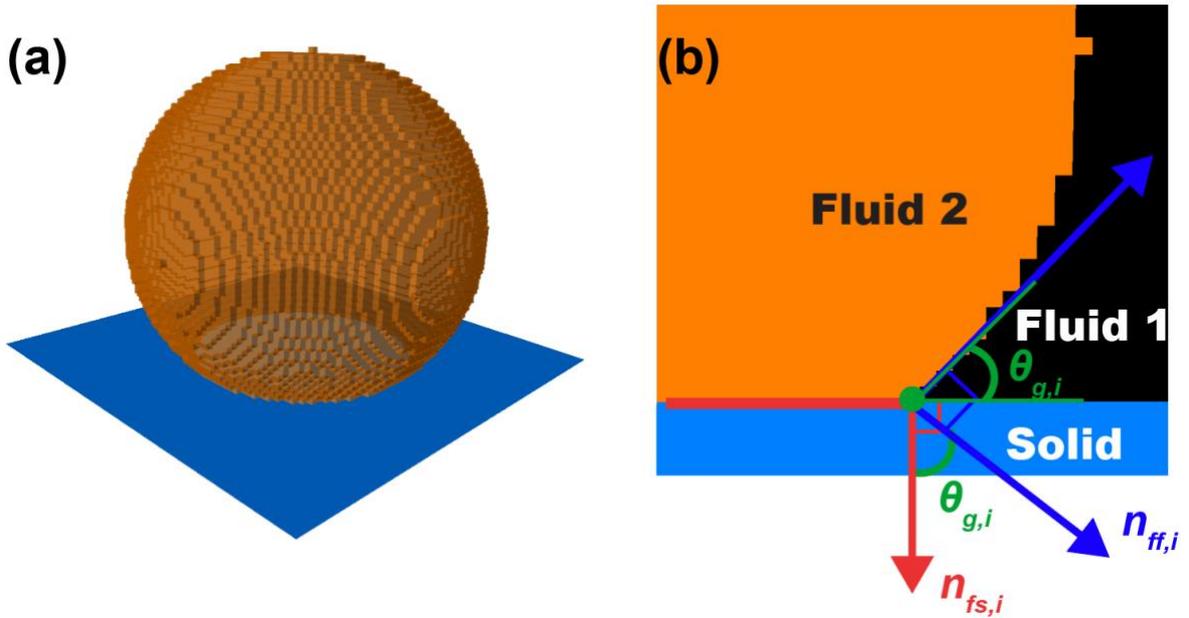

**Fig 3.** (a) Voxel-based 3D droplet model on a solid surface. (b) Cross-section showing Fluid 1 (dense, e.g., brine), Fluid 2 (light, e.g., oil/gas) and solid substrate. Normal vectors at the three-phase contact node: $\boldsymbol{n_{fs,i}}$ (fluid–solid) and $\boldsymbol{n_{ff,i}}$ (fluid-fluid) $\boldsymbol{\theta}$ between them is equivalent to $\boldsymbol{\theta_{g,i}}$.
10

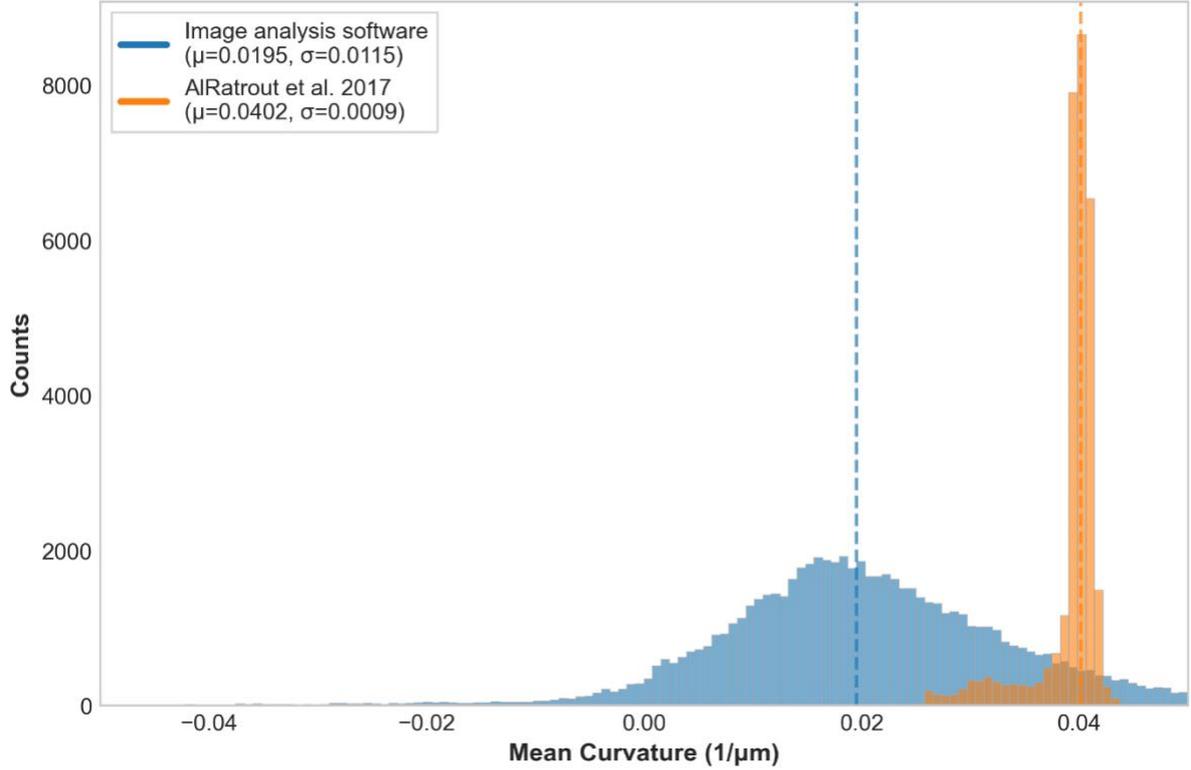

**Fig 4.** Mean curvature distribution of the fluid–fluid interface mesh for the synthetic droplet in Figure 1a (analytical $\kappa_m$ = 0.02 voxel$^{-1}$), comparing the results of commercial image analysis software (Avizo) and AlRatrout et al. (2017). AlRatrout et al.'s tool shows a higher mean, indicating interface shrinkage.

Three-phase contact loops, where solid, liquid, and gas phases converge, pose fundamental challenges for accurate segmentation. These challenges arise from multiple factors: (1) partial volume effects, where individual voxels contain contributions from all three phases, making precise boundary determination impossible (Schlüter et al., 2014); (2) the sub-voxel nature of contact loops, which are theoretically 1D features that cannot be precisely localized within the finite voxel resolution of laboratory imaging systems; and (3) poor phase contrast, particularly in oil-brine-rock systems where similar X-ray attenuation coefficients result in overlapping grayscale values (Singh et al., 2016). Additionally, high-resolution imaging can reveal surface roughness that introduces further geometric complexity at the contact loops (Andrew et al., 2014).

These factors combine to make capturing the true geometry of contact loops challenging from pore-scale images, as direct use of $n_{fs,i}$ and $n_{ff,i}$ would propagate segmentation errors into contact angle measurements. To address these limitations, we propose a novel approach that reconstructs normal vectors through extrapolation based on near-surface information—regions where segmentation is more reliable. This distance-based approximation yields robust results by reducing dependence on the problematic three-phase region and incorporating the broader interface surface mesh behavior. Furthermore, this approach enables proper identification and removal of extraneous nodes at the contact loops, enhancing the accuracy of contact angle estimation despite the inevitable segmentation uncertainties at three-phase boundaries.

From the voxelized 3D image containing three distinct labels (Section 2.2.2), Fig 5 outlines the methodological framework used to measure contact angles from the segmented voxel data. The three phases include solid, fluid 1 (the denser phase, e.g., brine), and fluid 2 (the lighter phase, e.g., gas). The algorithm proceeds in five primary steps: extract interfaces, identify three-phase contact loops, smooth both surfaces and contact loops, extrapolate $n_{fs,i}$ and $n_{ff,i}$, and contact angle calculation.

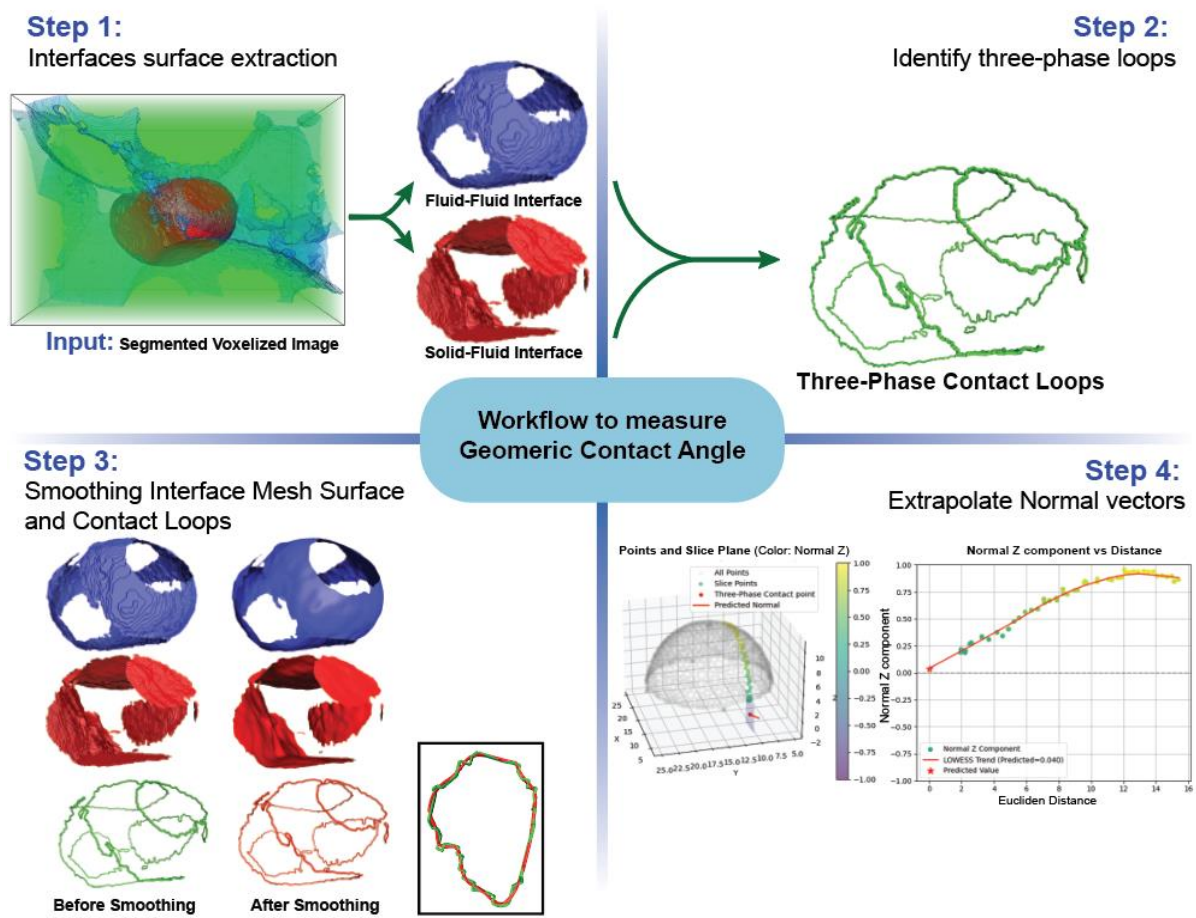

**Fig 5.** Workflow steps for measuring geometric contact angle from a segmented image.

**Interface Extraction**

To accurately reconstruct fluid-solid interfaces from voxel images, we converted the volumetric data into triangular mesh surfaces that precisely capture phase boundaries (Fig 6). We evaluated both triangular and quadrilateral mesh representations for surface extraction (Table 2). To assess mesh quality, we compared mean curvature measurements for spheres of varying radii (Table 2) where the sphere mean curvature ($\kappa_m$) is inversely proportional to its radius ($1/r$). Spherical geometries represent the terminal meniscus shapes commonly found at fluid-fluid interfaces in two-phase flow applications (Andrew et al., 2015). Under identical smoothing iterations, triangular meshes demonstrated superior performance accuracy.

We selected triangular meshes based on simplicity with available Python libraries like scikit-image and VTK provide efficient marching cubes implementations for triangular mesh generation, and geometric preservation that require lower smoothing iterations to achieve comparable surface quality compared to quadrilateral. As making minimal smoothing requirements essential for preserving interface geometry.

Surface generation employed the marching cubes algorithm (Lewiner et al., 2003; Lorensen and Cline, 1987), which approximates isosurfaces by converting voxel data into triangular elements for both the solid phase ($S_S$) and fluid 2 ($S_{F2}$).

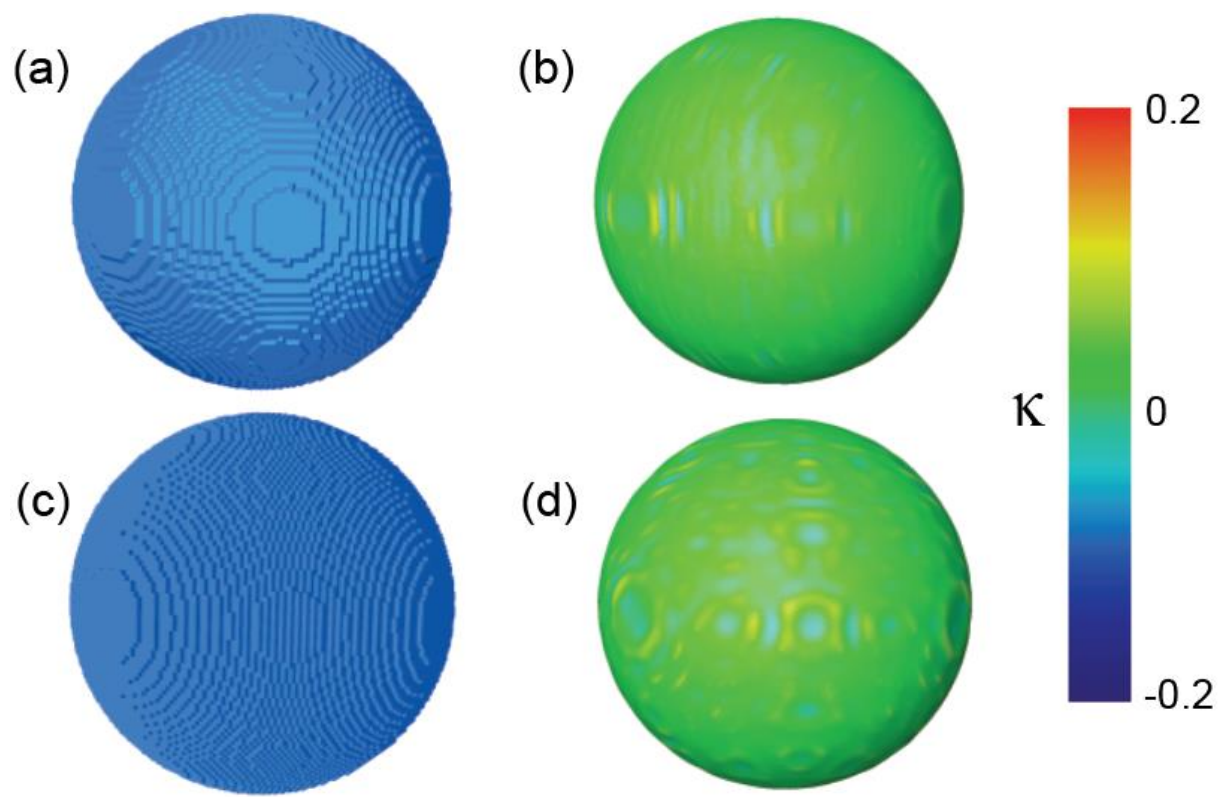

**Fig 6.** Comparison of two surface mesh types from the sphere case: (a) triangular mesh, (b) smoothed triangular mesh colored by mean curvature, (c) quadrilateral mesh, and (d) smoothed quadrilateral mesh.

**Table 2.** Comparison of curvature and mesh properties for quadrilateral and triangular surface meshes extracted from synthetic spheres of varying radii.

| Quadrilateral Mesh | Triangular Mesh | $\kappa_m = 1/r$ |
|---|---|---|

| Sphere Radius (voxels) | faces | vertices | $\kappa_m$(voxel$^{-1}$) | faces | nodes | $\kappa_m$(voxel$^{-1}$) | |
|---|---|---|---|---|---|---|---|
| 50 | 80530 | 40408 | **0.0224** | 80812 | 40549 | **0.0208** | 0.0200 |
| 28 | 26,010 | 13,080 | **0.0400** | 26,156 | 13,153 | **0.0373** | 0.0357 |
| 6 | 1154 | 596 | **0.1896** | 1188 | 613 | **0.1753** | 0.1667 |

Once the meshing is complete, the surfaces are categorized into two groups: fluid-solid (denoted by $I_{fs}$) and fluid-fluid (denoted by $I_{ff}$). Given the system structure, Fluid 1 is always adjacent to either the solid surface or fluid 2, in two-fluid-phase flow. To classify these surfaces, we first identify the shared faces ($f$) between the two isosurfaces, representing the fluid-solid interface ($I_{SF}$), The remaining faces are assigned to the fluid-fluid interface.

$$I_{SF} = \{f_i \in S_S \cap S_{F1}\}$$
$$I_{FF} = \{f_i \notin S_S \cap S_{F1}\}$$

---

**Algorithm 1: Interface Extraction from Voxelized Data**

**Input:**  $V \rightarrow \mathbf{R}^3$, *F2* (lighter fluid label), *S* (solid label), $V_{min}$ (cluster volume threshold)
**Output:**  $M_{FF}$ (fluid-fluid interface mesh), $M_{SF}$ (solid-fluid interface mesh)

**Phase 1: Volume Filtering**
1:      $\Omega \leftarrow$ Connected Components (*V* = *F2*)    // Identify fluid2 regions
2:      *V'* $\leftarrow$ *V*    // Create modified volume
3:      for $\omega_i \in \Omega$ do
        • $n_i \leftarrow$ CountVoxels($\omega_i$)    // Number of voxels in region
        • If $n_i < V_{min}$ then:    // Compare voxel count
           o *V'* $\leftarrow$ Remove Component (*V'*, $\omega_i$) // Update volume
        • end If
4:      end for

**Phase 2: Surface Generation**
5:      $(S_{F2}) \leftarrow$ Marching Cubes (*V'* = *F2*)    // Generate fluid2 surface
6:      $(S_S) \leftarrow$ Marching Cubes (*V'* = *S*)    // Generate solid surface

**Phase 3: Interface Classification**
7:      for each $f_i \in S_{F2}$ do
        • If Shares faces ($f_i \in S_S \cap S_{F2}$) then
           • $M_{SF} = f_i$    // Assign face to solid-fluid interface
        • Else
           • $M_{FF} = f_i$    // Assign face to fluid-fluid interface
        • end If
     end for

**Return** $M_{FF}, M_{SF}$

---

**Identifying Contact Loops**

Vertices that are shared between the fluid-solid ($I_{SF}$) and fluid-fluid ($I_{FF}$) interfaces represent the three-phase contact points. Let $V_{tp}$ be the set of vertices shared by both interfaces:

$$V_{tp} = V_{SF} \cap V_{FF}$$

From these three-phase vertices, we construct contact loops. Vertices that are adjacent to each other are connected to form contact loops. If the starting vertex is adjacent to the ending vertex, they form a contact loop (Fig 5). Denote the set of three-phase vertices as $V_{tp}$ and define $E$ as the set of edges:

For vertices $v_i, v_j \in V_{tp}$:

If $v_i$ and $v_j$ are adjacent, create an edge $e_{ij} \in E$

If $v_i$ is adjacent to $v_j$, and $v_j$ is adjacent to $v_0$ (starting vertex), connect $v_i$ to $v_0$ to complete loop.

Thus, a three-phase contact loop can be expressed as:

$$L = \{v_0, v_1, \ldots\ldots\ldots, v_n | (v_i, v_{i+1}) \in E, (v_n, v_0) \in E\}$$

---

**Algorithm 2: Define Three-phase Contact Loops**

**Input:** $M_{FF}, M_{SF}$
**Output:** $\mathcal{L} = (\ell_1, \ell_2, \ldots\ldots, \ell_m)$ (Set of three-phase contact loops)

**Phase 1: Three-Phase Vertices**
1:        $V_{tp} \leftarrow \{v_i : v_i \in V_{FF} \cap V_{SF}\}$      // Shared Vertices

**Phase 2: Define Edges**
2:        $E \leftarrow \{(v_i, v_j) : v_i, v_j \in V_{tp}, are\ adjucent\ (v_i, v_j)\}$

**Phase 3: Line and Loop Construction**
3:        $\mathcal{L} \leftarrow \Phi$      // Initialize line set
4:        For each $C$ in Connected Components $(V_{tp}, E)$ do
- Select $v_0$ from $C$
- $\ell \leftarrow$ Trace Path $(v_0, E)$      // Trace sequential path
- If Is Closed Loop $(\ell)$ then      // Check if path returns to $v_0$
  - $\ell$.type $\leftarrow$ "*loop*"
- else
  - $\ell$.type$\leftarrow$ "*line*"
- end if

       $\mathcal{L} \leftarrow \mathcal{L} \cup \{\ell\}$      // Add to set regardless of type
5:        End for
**Return** $\mathcal{L}$

---

**Smoothing**

To generate smooth and accurate representations of both contact loops and mesh surfaces, we employed two independent smoothing methods: one for the contact loop set, and another for mesh surfaces. For contact loops, we utilized B-spline fitting techniques (James et al., 2017), with knot adjustment to accurately capture geometric features while preserving original contact length. This approach provided continuous boundary representation with uniform node spacing, limiting outlier three-phase nodes crucial for accurate contact angle measurements.

For mesh surfaces, we applied Taubin smoothing (Taubin et al., 1996) to address the limitations of standard Laplacian smoothing. While Laplacian smoothing reduces noise by adjusting vertices to neighboring vertex averages, it causes volume shrinkage detrimental to geometric accuracy, as shown in Fig. 4. Taubin smoothing preserves both volume and surface features through a two-step process.

The vertex position adjustment in Taubin smoothing is calculated as $\Delta v_i = \frac{1}{|N(i)|} \sum_{v_j \in N(i)} (v_j - v_i)$, where $v_i$ and $v_j$ represent the current vertex and its neighboring vertices, respectively. Here, $N(i)$ is the set of neighboring vertices, and $|N(i)|$ is the number of these neighbors.. The first step adjusts the current vertex position as is $v_i^{(k+1/2)} = v_i^k + \lambda \Delta v_i^k$, followed by a second adjustment $v_i^{(k+1)} = v_i^{(k+1/2)} + \mu \Delta v_i^{(k+1/2)}$, where $\lambda$ and $\mu$ are smoothing parameter. The parameter $\lambda$ smooths of the mesh, while $\mu$ counteracts volume loss by inflating the mesh.

These complementary smoothing techniques ensure the final representation preserves physical interface features without over-smoothing critical structures, making it well-suited for contact angle analysis.

**Normal Vector and Contact angle**

For each contact node $v_i \in L$ along the contact loop, the local surface normals are extrapolated from nearby mesh faces. A local coordinate system is established at each node, with the secant direction computed from adjacent vertices $[v_{i-1}, v_{i+1}]$. serving as the primary axis. A plane perpendicular to secant line defines the search region for face selection.

Faces are selected $f_i \in I_{SF} \cup I_{FF}$ for extrapolation based on two criteria: (1) their center must lie within a slab of thickness $\pm d$ (typically 0.5 unit) perpendicular to the contact line, and (2) their Euclidean distance from $v_i$ must not exceed 12 units (the voxel length is 1 unit). This dual constraint ensures sufficient local data while excluding distant faces that could degrade extrapolation accuracy. A minimum of 20 faces is required for reliable normal estimation; nodes with insufficient neighboring faces are not considered.

The extrapolation method for $n_{ff,i}$ is selected based on the local interface geometry. For meniscus interfaces, polynomial regression is applied to faces within 3 units of the contact point, as polynomials become unstable when extrapolating beyond their fitting range. A second-order polynomial is typically sufficient for smooth surfaces. For interfaces with sharp features or high curvature variations, higher-order polynomials (3rd-4th degree) may be employed to capture the complex geometry.

Locally Weighted Scatterplot Smoothing (LOWESS) is reserved for nearly-flat or constant-curvature regions where polynomial fitting may overfit noise and for $n_{sf,i}$. However, LOWESS is always computed as a fallback option. A quality control mechanism ensures robust extrapolation: if the polynomial-extrapolated normal deviates more than 30° from the mean of the nearest measured normal, the algorithm reverts to the LOWESS result. This constraint prevents unrealistic extrapolations while maintaining the advantages of polynomial fitting in the near-field.

Following normal vector extrapolation, $\theta_{g,i}$ is calculated using equation (2). Post-processing ensures data quality through outlier detection, where contact angle values that deviate significantly from neighbouring points are identified and corrected using median absolute deviation analysis.

### 3.3. Generate spatially distributed contact angle map

We present a novel hierarchical approach for assigning contact angles throughout three-dimensional porous media based on sparse measurement data. The method employs phase-specific interpolation strategies that respect fluid topology and measurement uncertainty, with particular emphasis on the accurate treatment of primary drainage that is an invasion percolation process (Blunt, 2017; Wilkinson and Willemsen, 1983). The approach segments the pore space into distinct regions based on fluid occupancy and applies tailored interpolation methods to each region, resulting in physically consistent contact angle distributions. This method follows, (i) identification and treatment of uninvaded pore regions, (ii) object-based interpolation for the invaded phase (oil or gas), and (iii) distance-limited interpolation for remaining defending phase (brine) regions. The approach ensures physical consistency while accounting for measurement uncertainty.

The methodology operates on segmented 3D images where voxels are classified as solid ($\varphi = 2$), invading fluid ($\varphi = 1$), or defending fluid ($\varphi = 0$). The pore space, denoted $\Omega$, comprises all non-solid voxels: $\Omega = \{x \in \mathbb{R}^3 : \varphi(x) \in \{0, 1\}\}$. Contact angle measurements $\theta_{g,i}$, are provided at selected position $p_i$ along three-phase contact loops, where each measurement is associated with a specific line $\ell$ containing $v_i$ measurement points.

To account for strongly water-wet regions typically observed during primary drainage, uninvaded pore regions are identified by connected component analysis with 6-connectivity identifies pore regions $C_j$ containing only defending fluid ($\Phi_j = \{0\}$). These uninvaded regions, trapped by capillary barriers during primary drainage, are assigned $\theta_g = 30°$ to represent strongly water-wet conditions.

The invading fluid phase is segmented into individual objects using connected component analysis: this is done for computational memory efficiency. For each object, nearby contact lines are identified through 3-voxel morphological dilation of the object boundary. Each contact line's mean contact angle is weighted by its measurement count to reflect measurement confidence. Objects are then assigned contact angles using inverse distance weighting, where the interpolation considers the weighted mean angles from all associated contact lines. This approach ensures that objects receive spatially consistent contact angle values that reflect both the proximity to measurement locations and the reliability of those measurements, with lines containing more data points having greater influence on the final interpolated values.

Following the assignment of $\theta_g(x)$ to uninvaded regions and invading fluid objects, the remaining defending fluid voxels undergo interpolation using the same inverse distance weighting approach. These voxels, which exist outside the uninvaded pore regions, represent transition zones that have experienced wettability alteration due to proximity to three-phase contact lines. For each defending fluid voxel, the weighted mean contact angles from all measurement loops are considered. To prevent over-extrapolation into regions far from measurements, the interpolation is limited to voxels within a maximum distance of 20 voxels from any measurement point. This distance-constrained approach ensures that only defending fluid regions genuinely influenced by nearby wettability conditions receive interpolated values, while more distant regions remain unassigned, maintaining the physical validity of the contact angle field.

Finally, all assigned contact angle values are clipped to lie within the physically meaningful range of 1° to 180°, ensuring numerical stability and consistency with surface wetting physics.

## 4. Results and Discussion
### 4.1. Measurements Accuracy and Uncertainty

To evaluate the accuracy and robustness of our automated contact-angle algorithm, we designed three distinct tests—each chosen to probe a different source of error in voxelized datasets. First, we simulated a spherical droplet resting on a flat substrate (Fig 1a) and rendered it at five resolutions (sphere radii = 6, 10, 14, 28, and 50 voxels). Second, we positioned an identical droplet on a curved substrate (Fig 1b). Third, we analysed a droplet embedded in a segmented micro-CT image of Ketton Limestone (Fig 2), a configuration that lacks an analytical reference and therefore requires manual measurement.

In the flat- and natural-surface cases we benchmarked our extrapolation-based normal estimation against the currently most used automated open-source tool (AlRatrout et al., (2017)) (Fig 7 andFig 8), while in the curved-surface test only our method was applied (Fig 9). Both codes return a list of contact-angle values—one for every three-phase contact point in each contact loop—so the values shown in Fig 7a, 7b, and Fig 8 represent the arithmetic mean of those individual measurements.

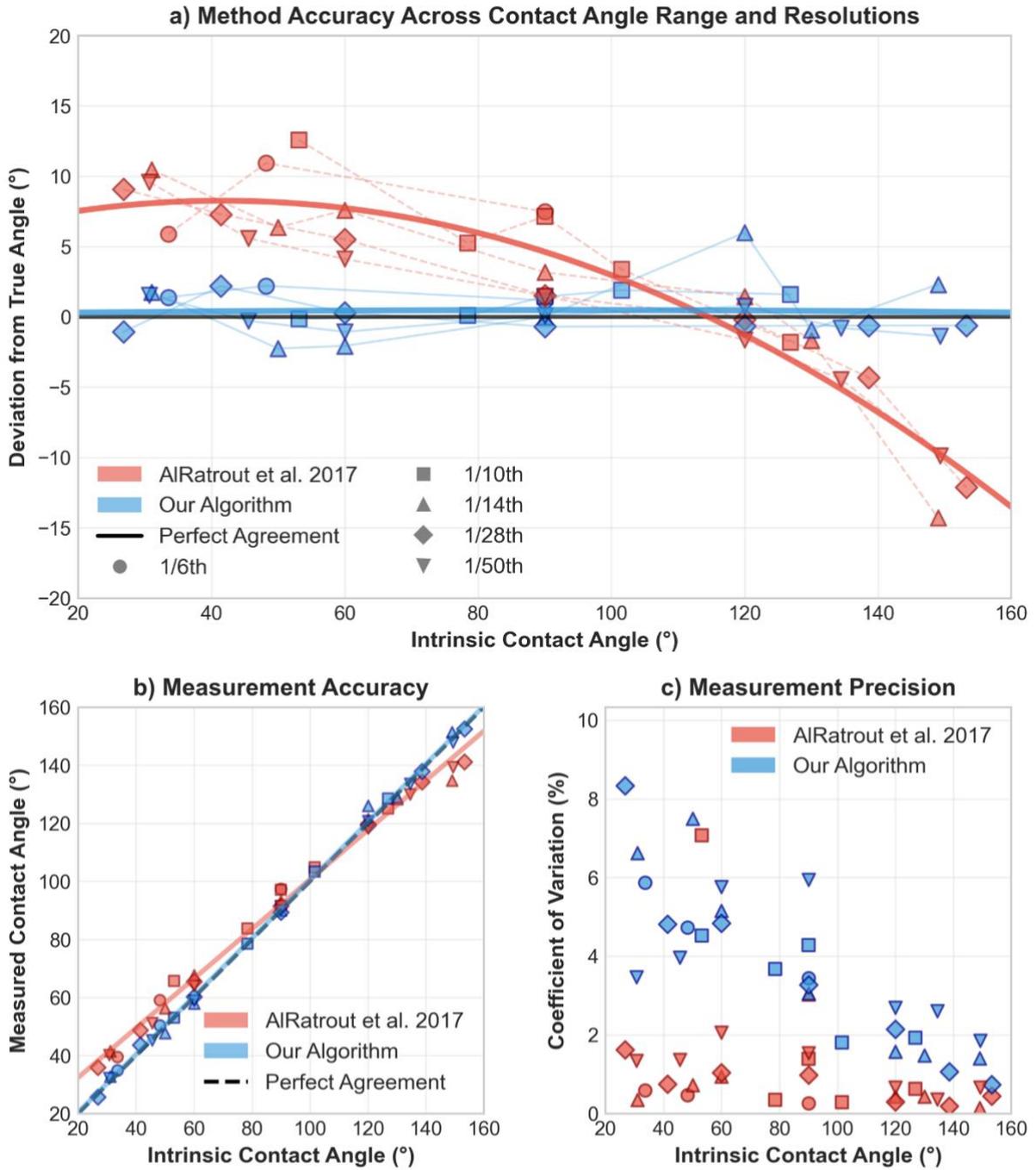

**Fig 7.** Flat surface case: performance comparison between our algorithm and AlRatrout et al. (2017).

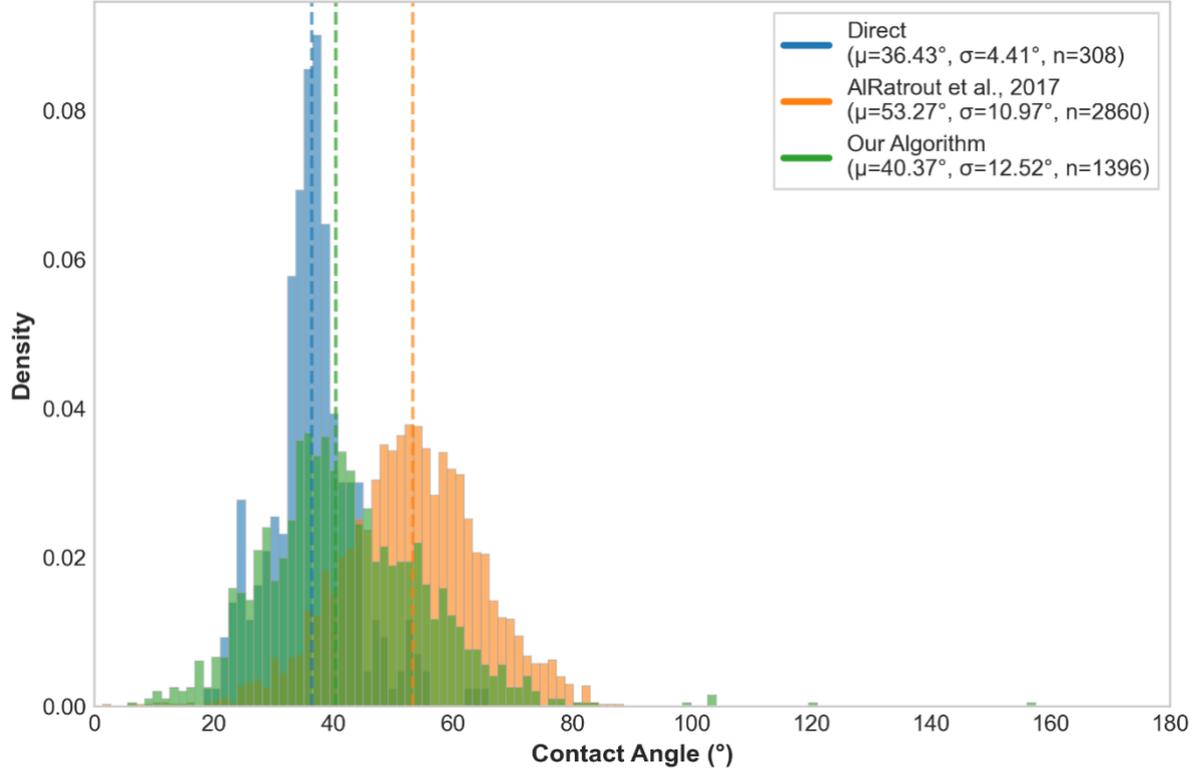

**Fig 8.** Contact angle histogram for Ketton limestone ganglion SSb, benchmarking two automated tools with manual direct measurements.

Across the entire contact-angle spectrum and at all tested resolutions, our algorithm performs consistently well, whereas the AlRatrout *et al.* code systematically overestimates low contact angles, underestimates high ones, and is markedly more sensitive to voxel resolution (Fig 7a). AlRatrout *et al.*s higher nominal precision—evident in Fig 7c—results from its volume-preserving curvature smoothing, which produces uniformly oriented normals for ideal geometries (e.g., a single-surface sphere). However, this smoothing can oversimplify real surfaces, causing a systematic shift in the mean contact angle. Our method, by contrast, estimates the surface normal independently at each contact point. This local approach minimizes the influence of segmentation artifacts near three-phase boundaries and preserves fine-scale geometric detail. This can also be observed in the Ketton limestone case (Fig 8), where our algorithm yields contact-angle distributions that are closer to direct manual measurements. Although the standard deviation is slightly higher than that produced by the AlRatrout *et al.* method, the mean value is more accurate. The AlRatrout *et al.* code, while producing narrower distributions, significantly overestimates the contact angle, which mischaracterizes the system as almost weakly water-wet rather than strongly water-wet.

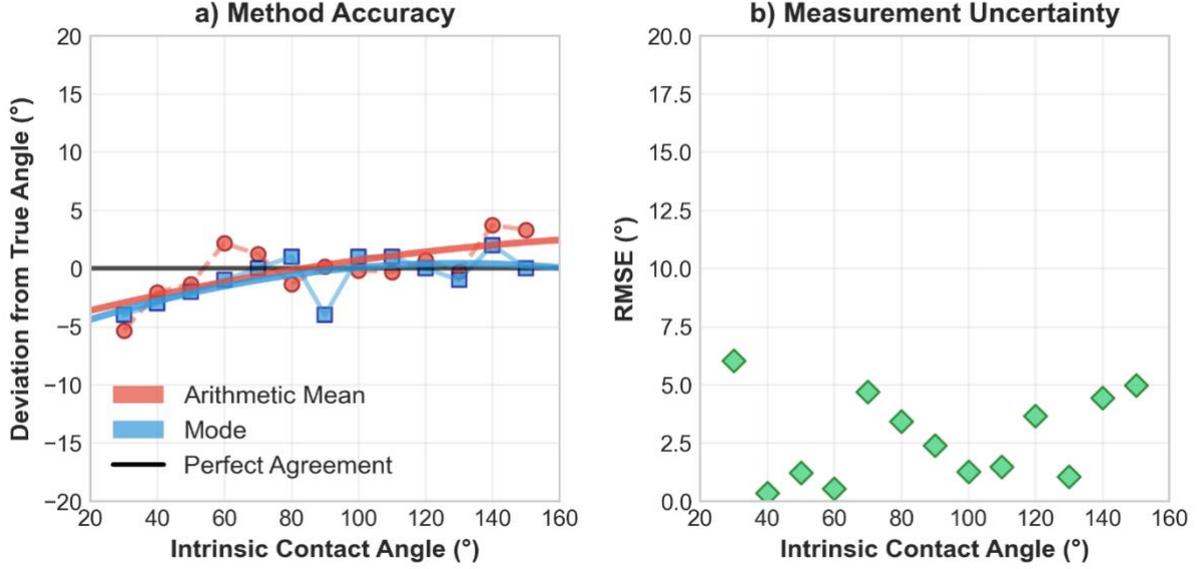

**Fig 9.** Algorithm performance on a curved surface case.

To assess estimator performance on curved surfaces, we compared contact-angle results computed using both the arithmetic mean and the mode (Fig 9a). In each approach, the computed angles remained within ±5° of the intrinsic angle; however, the modal estimator exhibited significantly less scatter. This reduction in variability reflects the mode's robustness against random, high-leverage outliers, an advantage also evident from the RMSE metric (Fig 9b). When comparing curved-surface to flat-surface results, slightly lower accuracy was observed for the curved surface. This is because, on flat surfaces, $n_{fs,i}$ remain consistent across the entire mesh, whereas on curved surfaces, these vectors vary spatially. On more complex geometries, increased deviations are anticipated due to surface roughness, noise, or segmentation artifacts, which introduce additional local heterogeneity. Consequently, we adopted the mean of the contact loops, for constructing spatially resolved contact-angle fields from voxel-wise data.

### 4.2. Tests on experiments images

Fig 10 presents the results from four reprocessed experimental images. We compared contact angle measurements obtained using three methods: our own code, the Altratorut *et al.* (2017), and an open-source geometric contact angle measurement tool developed by Wang (2024) in MATLAB. Wang's tool, originally designed to compare geometric and topological contact angles, identifies and analyzes individual phase clusters independently. However, it applies a maximum cluster size constraint, limiting analysis to ganglia with a radius of up to 64 voxels. Consequently, larger connected structures—such as ganglion SSa—could not be measured using this method.

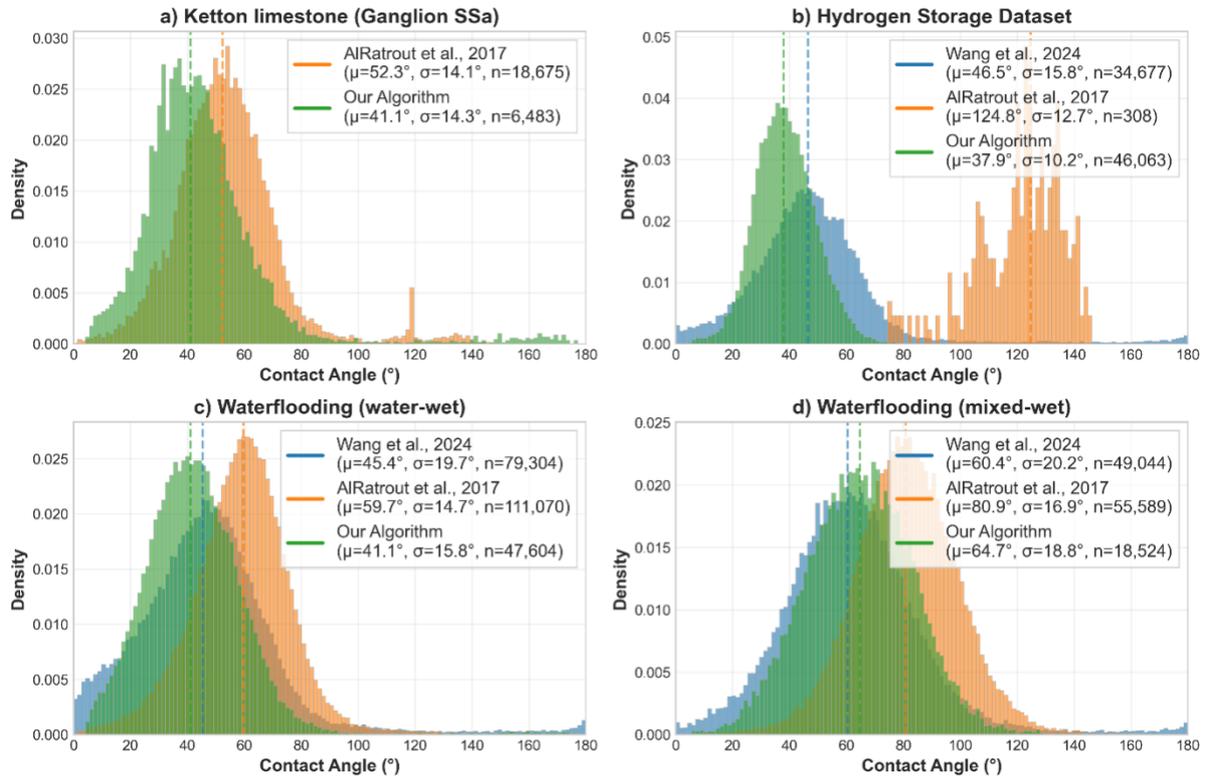

**Fig 10.** Histogram comparison of contact angle distributions from three automated geometric measurement tools.

In Fig 10a, our algorithm demonstrated robust performance with a mean contact angle of 41.1° ± 14.3°, closely matching the validated results of Scanziani et al. (2017), who reported 42.2° ± 14.6° from 4,025 measurements. This agreement builds confidence in measurement reliability for morphologically complex systems. Compared to AlRatrout *et al.,* our method yielded fewer valid data points due to contact loop smoothing and filtering of unreliable normal vector extrapolations (Section 3.2). This quality control ensures measurement reliability, as accuracy is more important than quantity when measuring contact angles from voxel image data.

The versatility of our approach is evident across diverse fluid systems and wettability conditions. For water-wet cases (Fig 10a-c), the algorithm consistently produces physically reasonable distributions centred at 37-41°, successfully handling both oil-brine and gas-brine systems where AlRatrout et al.'s method failed to generate measurements for the hydrogen storage dataset.

For the mixed-wet sample (Fig 10d), our measurements yield a mean contact angle of 64.7° ± 19.8°, which from a macroscopic averaging perspective indicates weakly water-wet conditions. However, this classification appears inconsistent with the presence of almost minimal interfaces—surfaces characterized by zero mean curvature and negative Gaussian curvature that typically form under intermediate-wet conditions ($\theta \approx 90°$). This discrepancy suggests that simple macroscopic characterization may misrepresent the complex wetting behavior. The high standard deviation reflects significant local wettability heterogeneity, as evident in Section 4.3, where some pores exhibit intermediate-wet conditions that explain the presence of minimal surface interfaces.

Comparative analysis reveals method-specific limitations: AlRatrout *et al.*'s approach generates smooth distributions with numerous data points but exhibits systematic overestimation due to curvature smoothing artifacts, while Wang *et al.*'s tool, designed primarily for cluster-based measurements, cannot process continuous phase distributions and thus has limited applicability for comprehensive contact angle characterization.

### 4.3. Spatially-distributed contact angles

We generated spatially distributed contact angle maps for waterflooding cases in both water-wet (Fig 11) and mixed-wet (Fig 12) systems to demonstrate the critical importance of local wetting information. The water-wet results reveal relatively homogeneous wetting behavior across all regions, with contact angles varying from strongly water-wet ($\theta < 45°$) to weakly water-wet ($50° < \theta < 70°$) conditions. While the displacement mechanism remains fundamentally the same throughout the sample (imbibition-dominated), the local variations in wetting strength affect displacement efficiency and capillary pressure gradients.

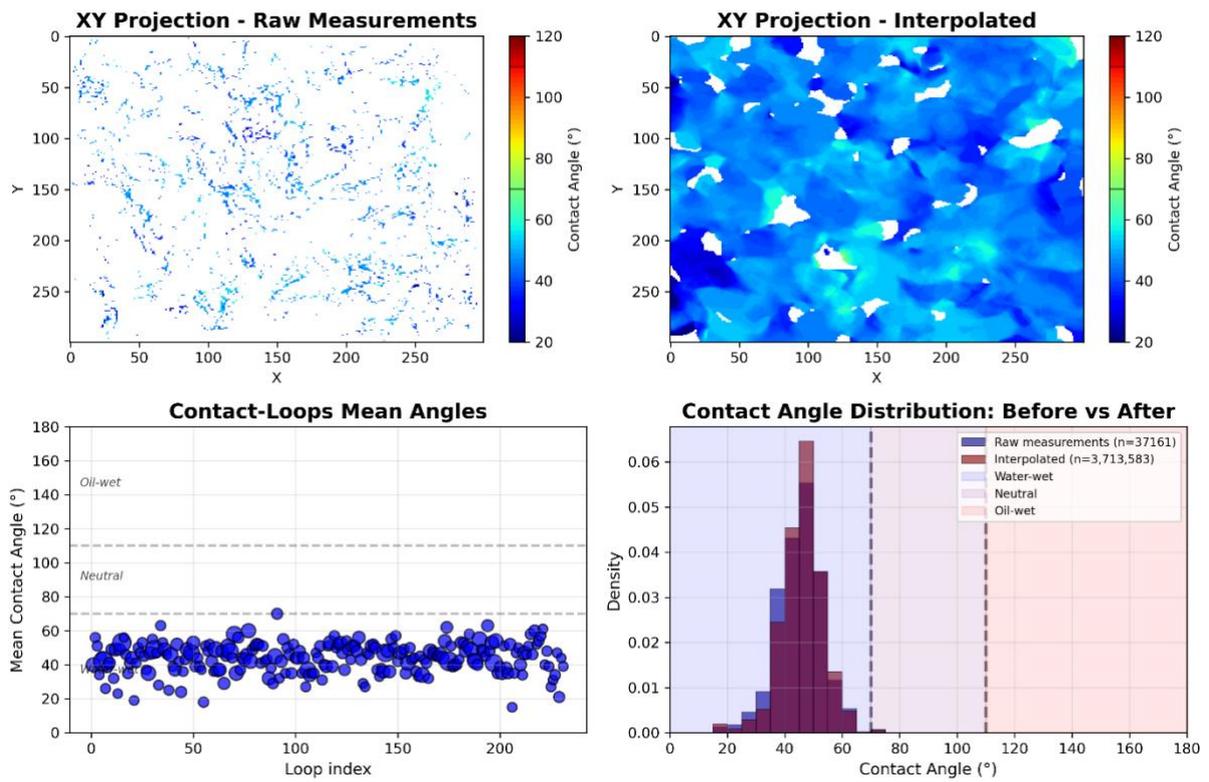

**Fig 11.** Spatial distribution of contact angles for a water-wet case. Top: XY projections showing contact angle distribution across all Z slices—raw measurements (left) and interpolated map (right). Bottom left: mean contact angle per contact loop. Bottom right: histogram comparing raw and interpolated distributions.

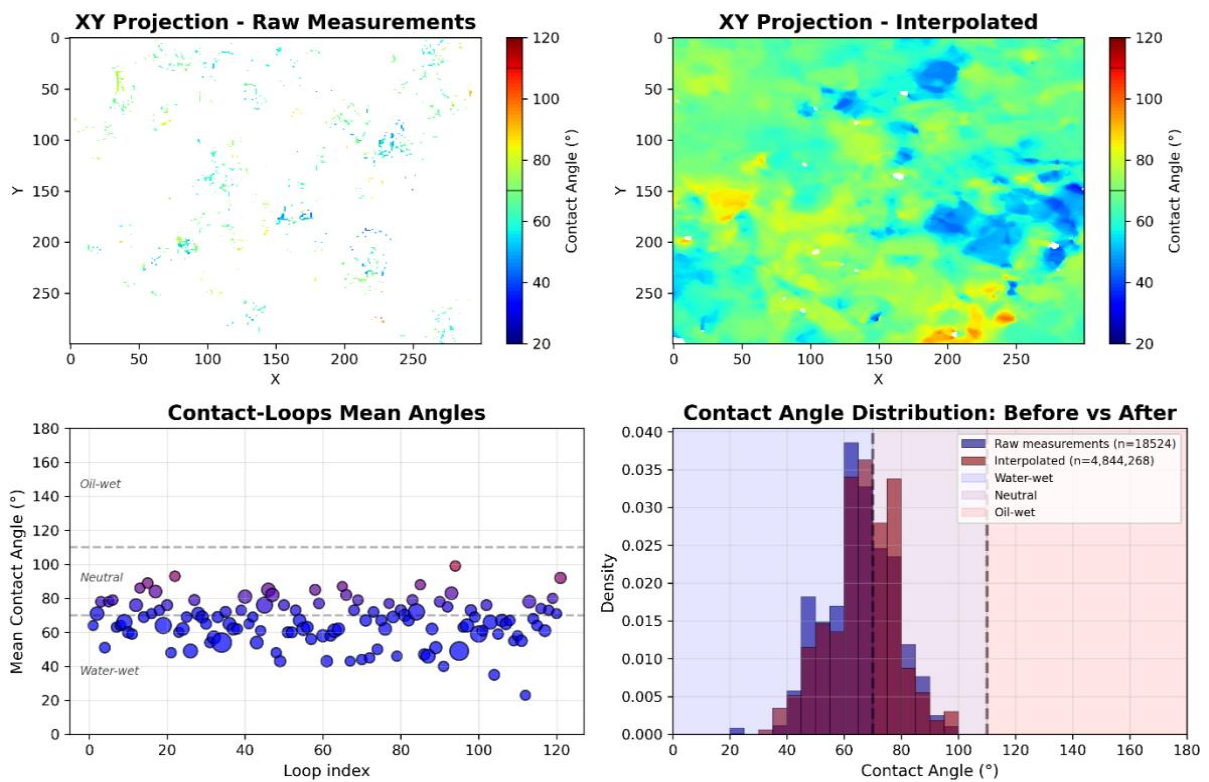

**Fig 12.** Spatial distribution of contact angles for a mixed-wet case. Top: XY projections showing contact angle distribution across all Z slices—raw measurements (left) and interpolated map (right). Bottom left: mean contact angle per contact loop. Bottom right: histogram comparing raw and interpolated distributions.

In contrast, the mixed-wet system exposes the severe limitations of relying solely on mean values for wettability characterization. Despite the mean contact angle of 64.7° suggesting uniformly weakly water-wet conditions, the spatial distribution reveals a dramatically heterogeneous system: 59.6% of the surface exhibits water-wet behavior ($\theta_g$ = 1-69°), while 40.4% displays intermediate-wet characteristics ($\theta_g$ = 70-110°), with some pores reaching contact angles of 100°. This heterogeneity is crucial for pore-scale simulations because fundamentally different displacement mechanisms operate in different regions—imbibition in water-wet pores versus forced displacement in intermediate-wet pores. The implications are profound: using only the mean contact angle would incorrectly predict uniform displacement behavior, potentially leading to significant errors in relative permeability predictions, residual saturation estimates, and breakthrough times. This spatial mapping capability is therefore essential for accurate multiphase flow modeling, particularly in carbonates and other systems where wettability alteration creates complex, heterogeneous wetting states that cannot be captured by averaged metrics alone.

## 5. Conclusions

This study addresses a fundamental challenge in characterizing wettability of porous materials, where recent advances have highlighted that spatially distributed wetting information, together with pore space geometry, controls displacement and trapping mechanisms in capillary-dominated flows. While assessing wettability from contact angles has long been questioned for accuracy and representativeness, our work demonstrates that geometric contact angle measurements, despite their theoretical simplicity, remain superior to other *in situ* methods, particularly in mixed-wet systems.

We present a novel surface reconstruction-based method for contact angle measurement near three-phase regions, designed to limit the artifacts of segmentation near three-phase contact loops. Our approach integrates both fluid–fluid and solid–fluid interface data using an extrapolation-driven geometric framework, enabling the construction of a sub-voxel surface mesh without relying on accurate three-phase voxel labeling. This segmentation-tolerant method significantly outperforms existing automated tools, particularly in cases with misclassified or noisy interfaces, and maintains high accuracy even under labeling uncertainty. Validated against analytical benchmarks, the algorithm demonstrates strong robustness and achieves pore-by-pore wettability measurements at an unprecedented spatial resolution. This allows for statistically meaningful, high-fidelity wettability characterization at the pore scale, previously unattainable with existing techniques.

Building on these measurements, we developed an automated algorithm to interpolate spatially distributed contact angles across entire pore networks, revealing the severe limitations of relying on averaged values. Our analysis of mixed-wet systems exemplifies this critical insight: while the mean contact angle of 64.7° suggests uniformly weakly water-wet conditions, spatial mapping reveals that 40% of pore space exhibit intermediate-wet behaviour ($\theta$ = 70-110°). This heterogeneity explains the presence of minimal surface interfaces and fundamentally different pore-filling mechanisms operating within the same sample—insights impossible to capture through averaged metrics alone.

While our extrapolation-based approach represents a significant advance, we acknowledge key limitations. The method requires clear interface boundaries for accurate measurements, making it sensitive to image resolution. Additionally, measurements from steady-state images may not capture the intrinsic contact angles required for dynamic pore-scale simulations, though corrections can be applied based on flow conditions.

To maximize impact and reproducibility, we provide a comprehensive open-source Python implementation, enabling the scientific community to leverage precise automated *in situ* wettability measurements. This contribution is particularly timely given the critical importance of accurate wettability characterization for $CO_2$ sequestration, hydrogen storage, and other sustainable energy applications where mixed-wettability conditions dominate system behavior. Moreover, these tools extend beyond their origins in geoengineering, offering valuable insights for any field where multiphase flow in porous materials plays a crucial role.

## 6. Declaration of competing interest

The authors declare that they have no known competing financial interests or personal relationships that could have appeared to influence the work reported in this paper.

## 7. Acknowledgment

The first author would like to thank Khalifa University of Science and Technology for providing the PhD scholarship that supported his research.

## 8. Data Availability

Code is available at https://github.com/ImperialCollegeLondon/geometricContactAngle.git.